%% file: main.tex
\documentclass{article}

\usepackage{arxiv}

\usepackage[utf8]{inputenc} 
\usepackage[T1]{fontenc}    
\usepackage{hyperref}       
\usepackage{url}            
\usepackage{booktabs}       
\usepackage{amsfonts}       
\usepackage{nicefrac}       
\usepackage{microtype}      
\usepackage{lipsum}

\usepackage{graphicx}

\usepackage{amsmath}

\usepackage[numbers]{natbib}

\usepackage{geometry}
\usepackage{float}

\usepackage{algorithm}
\usepackage{algpseudocode}

\usepackage[labelfont=bf]{caption}
\usepackage{tablefootnote}
\usepackage{multirow}      

\graphicspath{ {./images/} }

\title{Frequency domain kurtosis-based no-reference image quality assessment for bright-field microscopy images}

\author{
 Victor Augusto Alves Catanante \\
  Institute of Mathematics and Computer Science \\
  University of São Paulo, USP \\ Avenida Trabalhador são-carlense, 400, \\ 13566-590, São Carlos, SP, Brazil\\
  \texttt{vaugusto3692@gmail.com} \\
   \And
 Odemir Martinez Bruno \\
  São Carlos Institute of Physics \\
  University of São Paulo, USP \\
  PO Box 369, 13560-970, São Carlos, SP,\\
  \texttt{bruno@ifsc.usp.br} \\
  \And
 João do Espírito Santo Batista Neto \\
  Institute of Mathematics and Computer Science \\
  University of São Paulo, USP \\ Avenida Trabalhador são-carlense, 400,\\
  13566-590, São Carlos, SP, Brazil\\
  \texttt{jbatista@icmc.usp.br} \\
}

\begin{document}
\twocolumn[{%
  \begin{@twocolumnfalse}
    \maketitle
    \begin{abstract}
        In the last few years, image processing researchers spent a substantial amount of
        time and effort developing and perfecting image quality assessment algorithms. Bright-field microscopy, for example, produces images whose quality is a bottleneck for consistent evaluation. For instance, when a stack of images of a specimen is acquired in different focal plane configurations, there will be a set of blurred or partially blurred elements in it, impairing proper evaluation. This work aims to provide an image quality assessment metric, without the presence of a reference image for comparison, to detect the blurred and sharp images among the whole set of the stack, and elect the sharpest ones for a further fusion process. The correlation of the results with subjective labeling of the image sets showed that the proposed metric offers reliable identification of the eligible images for fusion and suggests the application in other real-world problems. 
    \end{abstract}
  \end{@twocolumnfalse}
}]


\section{Introduction}
\input{sections/introduction}

\section{Related work}
\label{sec:related-work}
\input{sections/related-work}

\section{Fourier Transform}
\label{sec:fourier_transform}
\input{sections/fourier-transform}

\section{Proposed method}
\label{sec:proposed_method}
\input{sections/proposed-method}

\section{Experiments}
\label{sec:experiments}
\input{sections/experiments}

\section{Results and Discussion}
\label{sec:results_and_discussion}
\input{sections/results-and-discussion}

\section{Conclusions}
\label{sec:conclusions}
\input{sections/conclusions}

\section*{Acknowledgments}

The authors would like to thank the Scientific Computing Group (SCG) from the São Carlos Institute of Physics for the bright-field microscopy lab. This work was supported by the Brazilian research agency Conselho Nacional de Desenvolvimento Científico e Tecnológico - Brasil (CNPq), process 132795/2018-3. This study was financed in part by the Coordenação de Aperfeiçoamento de Pessoal de Nível Superior - Brasil (CAPES) - Finance Code 001.

\bibliographystyle{plainnat}  
\bibliography{main} 

\end{document}

%% file: sections/introduction.tex
Images are present in almost every practical and theoretical field of human knowledge nowadays. From health and life sciences to public security systems, there are computational applications that offer some useful service employing image processing. As a result, assessing image quality poses as an important task among those applications for which several techniques are being developed, evolved and deployed. According to \citet{tang2019feature}, the image quality assessment (IQA) methods are distributed between the subjective assessment and objective assessment categories. The former is based on a well-defined test environment for random observers to label images and provide the final mean opinion scores (MOS), while the latter is based on the use of strategies such as statistical modeling, machine learning, spatial or spectral image features and so on. It is evident that subjective IQA is demanding; consequently, objective methods are preferred to conduct IQA.

According to \citet{wang2004image}, there are three classes of objective image quality metrics that relate to the existence of a no-distortion image (or with a negligible amount of it) for comparison purposes. The \emph{full-reference} (FR-IQA) methods assume that the reference image is available, while \emph{reduced-reference} (RR-IQA) methods employ a representation of the reference image, such as a set of extracted features. Finally, the \emph{no-reference} (NR-IQA) methods, also known as ``blind'', are those which do not employ a reference image. 

Microscopy is an important example of an image processing application that benefits from IQA methods. There are several different types of microscopy images: they may be acquired with lasers, transmitted or reflected light, measurements of atomic force responses, the fluorescence of chemical compounds and so on. Each microscopy technique has an inherent kind of degradation that affects the acquired images or spectra, e.g. the Raman confocal microspectroscopy suffers from the interference of cosmic rays, which yields unexpected peaks in the spectrum. 

In this work, we focus on the bright-field microscopy, especially the transmitted and reflected light microscopy. There are many applications that employ these techniques, such as biological analysis of cells, material sciences and plant leaf histology. In the scope of IQA, the main degradations perceived are noise and defocus blur, caused by the natural characteristics of the imaging device and optical properties. A common way of acquiring light microscopy images is the z-stack, where the focal plane is constantly changed. Considering that every object has an irregular surface, each focal plane will provide a sharp view of different regions of it. Indeed, a substantially sharp image may be built from all the sharp regions of each image from the dataset.

This work introduces a new NR-IQA method for light microscopy images in order to identify the sharpest images within a dataset, and also propose a benchmark dataset for IQA methods for this type of image. We compare the performance of different methods and analyze their behavior when subjected to this scope.

This paper is organized as follows: section \ref{sec:related-work} presents relevant research about the subject and situates our work in the scope of quality assessment. Section \ref{sec:fourier_transform} summarizes the mathematical background that guided the development of our method. Section \ref{sec:proposed_method} provides details about the method and its steps. Section \ref{sec:experiments} points out the settings for experiments, such as the image datasets we have proposed and the employed quantitative evaluation methods. Section \ref{sec:results_and_discussion} compares the results of the proposed method and other IQA methods when submitted to tests with our datasets, provides an analysis of the computational performance of all the tests and discusses how the nature of bright-field microscopy images may influence the performance of an IQA technique. Finally, section \ref{sec:conclusions} summarizes our work, indicates future work guidelines and describes possible applications of our method.

%% file: sections/related-work.tex
Because the blind measurement of distortions (NR-IQA) in images is not a trivial task, several attempts to develop methods to solve it in the past few decades have been made. There are two categories of 
NR-IQA methods: a) general-purpose, which
evaluate various types of distortions simultaneously
and b) distortion-specific, which rely on \emph{a
priori} knowledge of the distortion type \cite{tang2019feature}. This work presents a distortion-specific NR-IQA method that aims to quantify the blur information and, by means of statistical analysis, distinguish sharp and blurry elements in image datasets. This process can bring many practical benefits. Especially, in the microscopy field, we pinpoint the image fusion process to produce a distortion-free image.

A Fourier Transform-based algorithm was proposed by \citet{kanjar2013image} to compute an image quality measure of blurred images. This approach quantifies sharpness by computing the fraction of high-frequency components which in blurred image occurs at a smaller rate if compared with a sharper one.

\citet{ferzli2009noreference} proposed a blur metric that employs a probability summation model, taken from a psychometric function of just noticeable blur, i.e. a numerical representation of the human perception about contrast and blur.

\citet{narvekar2011noreference} improved the work of \citet{ferzli2009noreference}. Their work also looks at the information which is below the just noticeable blur limit, i.e. the probability of not detecting blur on an edge; the image quality index is the sum of those probabilities for the whole image. Thus, sharpness is in direct proportion to the CPBD value.

Spatial and spectral information from the image was also used to create the $S_{3}$ image
quality index \cite{vu2012s3}. First, the images are converted to the grayscale color space. For the spectral quality score, named $S_{1}$, the authors apply the two-dimensional Discrete Fourier Transform in images with significant contrast and use the slope of the magnitude spectrum in a sigmoid function to calculate it. In the computation of the $S_{2}$ index (related to the spatial information), they find the Total Variation (TV) index among 8-neighborhoods of pixels, then find the maximum of all neighborhoods. The $S_{3}$ metric is then the association of $S_{1}$ and $S_{2}$.

Still based on the TV approach, \citet{bahrami2014fast} propose the Maximum 
Local Variation (MLV) image quality index. The metric computes the maximum of grayscale level differences around 8-neighborhoods of all pixels, resulting in the MLV map. The map is weighted to emulate the human visual perception and later represented by a Generalized Gaussian Distribution; the quality score is its standard deviation, which increases as sharpness increases.

The perceptual blur metric presented by \citet{marziliano2002noreference} processes information within the edges of the image. The vertical edges obtained from an edge detector are scanned to find pixels that correspond to edges; for each pixel, the edge width is computed, and the final blur measure consists of an average of all widths.

The proposed method falls into the second group of NR-IQA methods since it is built to deal with the defocus blur stemming from acquiring images in different focal planes. In the scope of this work, the image quality metric should be capable of showing a subset of all available images that are eligible for the fusion process, which will search for the sharpest regions of each image of the subset and merge them in order to produce a better quality image.

%% file: sections/fourier-transform.tex
The Fourier Transform is a widely employed mathematical analysis tool, designed for studying the representation of a function with sums of simple trigonometric functions and their coefficients, i.e. a sum of Fourier series. In other words, it is a tool to represent any domain by their frequencies.

According to \citet{bracewell1978fourier}, the forward (\ref{eqn:1d_ft}) and the inverse (\ref{eqn:1d_ift}) Fourier Transforms for a one-variable function \linebreak $f(x)$ are respectively given by

\begin{equation}
    \label{eqn:1d_ft}
    \hat{f}(s) = \int_{-\infty}^{\infty}f(x)e^{-i2 \pi xs}dx
\end{equation}

\begin{equation}
    \label{eqn:1d_ift}
    f(x) = \int_{-\infty}^{\infty}\hat{f}(s)e^{i2 \pi xs}ds,
\end{equation}

\noindent where $\hat{f}(s)$ is the representation of the $f(x)$ function in the frequency domain and $e^{-i2 \pi xs}$ and $e^{i2 \pi xs}$ are the \emph{kernels} for the forward and inverse transformations, respectively. This formulation provides a representation of the function in \linebreak terms of its frequencies, considering that it was before in any one-dimensional domain, e.g. the time domain.

In the scope of image processing, images are described as functions of two variables. The two-dimensional Fourier Transform may be used to extract implicit frequency information from the image by transforming it from the spatial domain into the Fourier domain. Digital images are represented by matrices of pixels, and consequently, the Discrete Fourier Transform (DFT) may be applied to an image. The DFT as stated by \cite{bracewell1978fourier} is described here as 

\begin{equation}
    \label{eqn:2D_DFT}
    F(\mu,\nu) = M^{-1}N^{-1}
            \sum_{x = 0}^{M-1}
            \sum_{y = 0}^{N-1}
            f(x,y)
            e^{-i2 \pi (\frac{\mu x}{M} + \frac{\nu y}{N})},
\end{equation}

\noindent where $F$ is the DFT of the image, represented by the $f$ function. The variables $\mu$ and $\nu$ may be interpreted as frequencies taken in the $x$ and $y$ directions, respectively.

Although the main interest of the DFT in this work is the two-dimensional version, it may be applied to as many dimensions as needed, i.e. to each specified dimension of the data. 

\subsection{Image Degradation and the Fourier Spectrum}

Dirac Delta functions are generalizations of impulses, i.e. infinitely large values within an infinitely small time interval. The continuous Dirac Delta may be written as

\begin{equation}
\label{eqn:dirac_delta_function}
\delta^{2}(x,y)= 
\begin{cases}
    \infty, & \text{if } x^{2} + y^{2} =0\\
    0, & \text{if } x^{2} + y^{2} \neq 0
\end{cases}
\end{equation}

\noindent subject to the constraint

\begin{equation}
    \label{eqn:dirac_delta_constraint}
    \int_{-\infty}^{\infty}
    \int_{-\infty}^{\infty}
    \delta^{2}(x,y)dxdy = 1,
\end{equation}

\noindent The discrete version of the Dirac Delta function consists of an infinite sum instead of the integral. It is also useful to define the concept of Point Spread Function (PSF) of an imaging device: it is the representation of a two-dimensional impulse from a light source, which forms a point-shaped white object. The PSF is an extended blob in an image that describes a single point object, which can be mathematically described as a low-pass kernel.

Digital images are created using imaging devices, e.g. an optical microscope. Those are capable of capturing information from the continuous scene and create a discrete representation, by means of sampling and quantization. The process of digital image formation can be represented by

\begin{equation}
    \label{eqn:image_formation}
    g(x,y) = f(x,y) \ast h(x,y) + \eta(x,y),
\end{equation}

\noindent where $f(x,y)$ is the original image (without any degradation), $g(x,y)$ is the image after all the degradation processes, $h(x,y)$ is the PSF of the imaging device and $\eta(x,y)$ is a function which describes the noise conditions in which the image was taken. The symbol $\ast$ denotes the convolution operation, which is the process of flipping a filter mask by $180^\circ$, moving it along the image and computing the sum of the products at each location \cite{gonzalez2006digital}. The convolution operation in equation \ref{eqn:image_formation} is defined by 

\begin{equation}
    \label{eqn:2d_discrete_convolution}
    f(x,y) \ast h(x,y) = 
    \sum_{m=-a}^{a}
    \sum_{n=-b}^{b}
    f(m,n)h(x-m,y-n),
\end{equation}

\noindent where $a = (m-1)/2$ and $b = (n-1)/2$, given that the function $h(x,y)$ is considered to be a two-dimensional filter of size $m \times n$.

\begin{figure}[ht]
	\centering
		\includegraphics[scale=0.5]{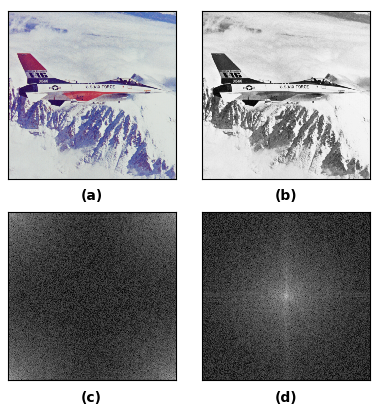}
	\caption{Original image \textbf{(a)}, luminance grayscale converted image \textbf{(b)}, unshifted Fourier spectrum of the grayscale image \textbf{(c)} and shifted Fourier spectrum of the grayscale image \textbf{(d)}.}
	\label{fig:airplane_fft_shift}
\end{figure}

The Fourier Spectrum is the amount of each frequency component among a discrete range of them, described in the form of a distribution. After the transformation, the resulting Fourier spectrum of the image consists of a matrix with complex coefficients and zeros on each of its four corners. Usually, the applications require a shift between the first and third quadrants, and also the second and the fourth quadrants, to the center of the matrix. The unshifted and shifted Fourier transforms of an grayscale airplane test image are shown in Figures \ref{fig:airplane_fft_shift}.(c) and \ref{fig:airplane_fft_shift}.(d), respectively. 

\begin{figure}[ht]
	\centering
		\includegraphics[scale=0.43]{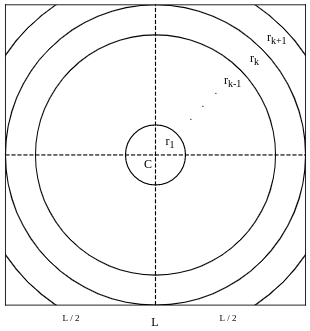}
	\caption{Frequency bands as rings of radius $\{r_{i}: i\in\mathbb{N}^{*}\}$ drawn over the 2D spectrum.}
	\label{fig:spectrum_bands}
\end{figure}

The frequency profile can be efficiently computed by \linebreak zero-padding the grayscale image before the transform so that the resulting image is a square matrix with power-of-two dimensions. The Fast Fourier Transform (FFT) is a \emph{divide and conquer} algorithm to reduce the computational complexity of the DFT from $\mathcal{O}(n^{2})$ to $\mathcal{O}(n\log{}n)$, which needs a power-of-two input sample size \cite{gonzalez2006digital}. Subsequently, each quadrant of the resulting matrix with the coefficients was shifted in order to achieve the same configuration of Figure \ref{fig:airplane_fft_shift}.(d). Let the matrix of DFT coefficients be represented as a square of side $L = \max{(m,n)}$, $k = L / 2$ be the maximum radius value for circles within the square and $C = (k,k)$ be the center of the infinite set of concentric circles inscribed in the square. Each circle represents a mask above the spectrum and stands for a frequency band, which starts as zero in the center of the matrix and increases together with the radius, as shown in Figure \ref{fig:spectrum_bands}.

Therefore, complex coefficients within the concentric circles with small radius values tend to have more energy than ones with radius values closer to k. Circles with increasing radius size, i.e. $\{r_{i},...:i \in \{1,2,...,k\}\}$, will cover all the frequency information from the image. The circles of radii $\{r_{i},...:i \in \{k+1,k+2,...\}\}$ comprise a very small area of the spectrum, and therefore may not be considered. Blurred images, for example, exhibit more low-frequency components than high-frequency ones. This is because blur may be understood as a filtering procedure with a low-pass kernel.

%% file: sections/proposed-method.tex
\begin{figure*}
  \centering
  \includegraphics[width=0.99\textwidth,height=4cm]{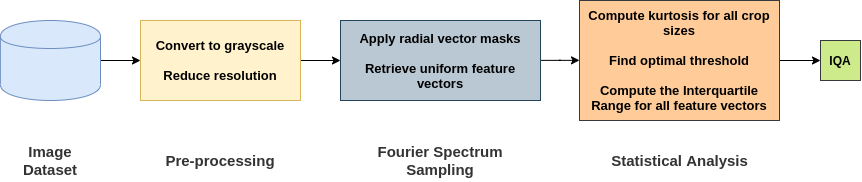}
  \caption{Pipeline of processing steps of the method.}
  \label{fig:pipeline}
\end{figure*}

We propose a new method for image quality assessment based on a sampling process of the Fourier spectrum and posterior analysis of the coefficients as a probability distribution by means of summary and descriptive statistics. Figure \ref{fig:pipeline} shows a diagram of the proposed method.

\subsection{Pre-processing}
\label{subsec:pre_processing}

The color space is a crucial feature
for the application since it
synthesizes the information from the image in a one-dimensional element. 
In our method, the image undergoes a grayscale conversion with the luminance method \cite{ponti2016image} - a 
linear combination of the three channels of an image from a trichromatic space such as RGB, as shown in the matrix equation given by

\begin{equation}
    \label{eqn:luminance}
    I_{luminance} = 0.299R + 0.587G + 0.114B,
\end{equation}

\noindent where $I_{luminance}$ is the matrix that represents the grayscale converted image, $R$, $G$ and $B$ represent the matrices of the red, green and blue channels, respectively.

Next, the resulting grayscale image resolution is reduced. The images in our dataset are of high resolution (2560 $\times$ 1920), rendering the process unfeasible. The resizing procedure consists of a bilinear interpolation, which uses the four nearest neighbors to estimate the intensity at a given location \cite{gonzalez2006digital}. It can be described by

\begin{equation}
    \label{eqn:bilinear_interpolation}
    I_{resized}(x,y) = ax + by + cxy + d.
\end{equation}

The pre-processing step ends with image enhancement. Microscopy images are, in general, acquired in different illumination conditions as a result of the operator's setup, focus adjustment by moving the objective or the stage and even physical properties of the microscope itself like transmitted or reflected light. To overcome this and deliver a uniform image for the Fourier Transform, a mapping function is used to perform contrast enhancement. The chosen algorithm for this task is the Contrast Limited Adaptive Histogram Equalization (CLAHE). It consists of computing several local histograms and distributing the gray levels along the regions from where the histograms were computed. The distribution, in this case, is done within a threshold to keep homogeneous areas and to reduce the noise amplification that occurs in a standard Adaptive Histogram Equalization \cite{zuiderveld1994constrast}. 

\subsection{Fourier Spectrum Sampling}
\label{subsec:fourier_spectrum_sampling}

As described in section \ref{sec:fourier_transform}, concentric circles along the shifted Fourier spectrum may be drawn in order to retrieve information from each frequency band. This approach is rather theoretical since the number of masks that may be applied to the spectra is finite. Taking into account the pixel resolution of the input images, it makes sense to sample the information, otherwise the computational complexity and running times
of the algorithm for one image alone would be impractical.

To comprise as much information about each frequency band as possible, we propose to sample the spectrum by means of radial lines as masks, i.e. white antialiased lines are drawn over a matrix of zeros, which are then element-wise multiplied by the spectrum. The lines are created from the $(x_{c},y_{c})$ center of the spectrum to points in an approximate radial position, which is calculated by

\begin{equation}
\label{eqn:points_on_radii}
P(x,y) = 
    (
    x_{c} r_{j} \cos{a_{j}}, 
    y_{c} r_{j} \sin{a_{j}}
    )
\end{equation}

\noindent with the set of angles $\{a_{j}\}$ in the radian form, computed as

\begin{align}
\label{eqn:angles}
\left\{
a_{j} : a_{j} = 
\frac{j \pi}{180}
\right\}
&&  j = \{0,5,...,100\}.
\end{align}

\noindent The outcome of equation
\ref{eqn:points_on_radii} is a
floating-point ordered pair, which is
rounded to the nearest integer value. The
antialiasing is achieved with a Gaussian filtering process. One example of all the generated lines is shown in figure \ref{fig:radial_masks}:

\begin{figure}
	\centering
		\includegraphics[
		scale=0.275]
		{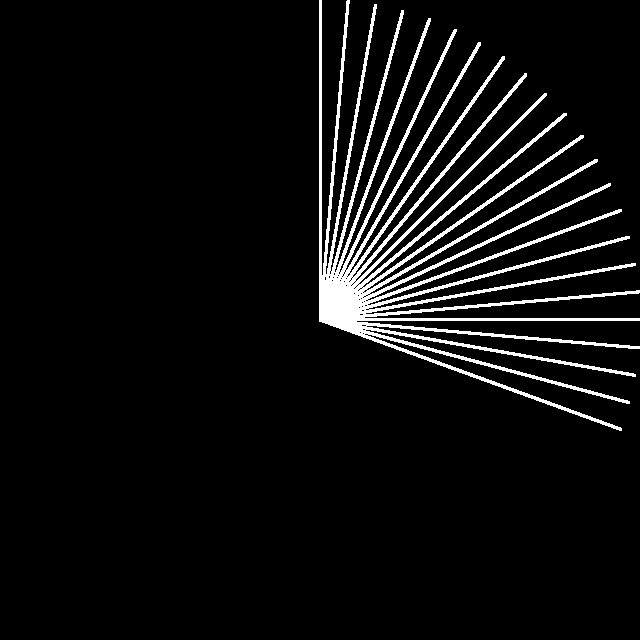}
	\caption{Final mask of radial lines.}
	\label{fig:radial_masks}
\end{figure}

\noindent After the element-wise multiplication, the radial vectors result in arrays of complex coefficients that represent samples of the frequency profile of the image. The radial lines in Figure  \ref{fig:radial_masks} have different lengths, hence the length of each array is not the same, even with antialiasing. Therefore, the shortest length value among all the vectors is taken as a limit. All vectors go through an element-wise average, which results in a one-dimensional vector as a descriptor of the frequency spectrum. To obtain a uniform feature vector, we set the smallest vector size among all of them and discard them. Therefore, if $L$ is the size of the zero-padded square image, the dimension of the final feature vector is about $1 / 2L$.

\subsection{Statistical Analysis}

As a result of the feature extraction process described in sections \ref{subsec:pre_processing} and \ref{subsec:fourier_spectrum_sampling}, we have a low-dimensional and concise representation of the image that captures blur information. We propose a set of steps to analyze the dataset which relies on statistical tools and the mathematical properties of the Fourier spectrum. 

Note that each feature vector is a distribution, with values in the range $[0,\infty)$. To use them as a probability distribution function suitable for techniques such as descriptive statistics and Bayesian inference they must be mapped onto the probability space $[0,1]$.  Hence, we apply a linear operator $T : \ell^{2}(\mathbb{Z}^{2}) \rightarrow \ell^{2}(\mathbb{Z}^{2})$, written as

\begin{align}
\label{eqn:probability_operator}
x_{i} = \frac{x_{i}}{\sum_{j=0}^{n-1}x_{j}}
&&  i = \{0,1,...,n-1\},
\end{align}

\noindent where each $x_{i}$ is a value of the descriptor which will be mapped onto a probability. 

Information embedded in the low-frequency components of the descriptor, which corresponds to the Dirac delta distribution within the point spread function of the imaging system, should be discarded. However, it should be done with caution, so that the remaining information is enough to represent the blur profile of the image. In order to properly discard the Dirac delta components from each descriptor, we propose to find an optimal threshold that allows the data to be ``cropped'', i.e. a subset of it will be taken as the final representation of the blur profile. The threshold is chosen to maximize the difference between the maximum and the minimum among a set of kurtosis values that represent each descriptor.

The optimal threshold is computed as follows. We start with a crop size equal zero by computing the kurtosis of the entire set $\{x_{1},x_{2},...,x_{n}\}$ of all descriptors. The crop size is then incremented by 1, yielding the subset $\{x_{2},x_{3},...,x_{n}\}$. This process is repeated until the kurtosis of all crop sizes
is computed. The kurtosis is one of the probability
distribution shape statistics: a measure of how large the "tail" of the distribution is, such that smaller absolute values indicate that the distribution tends to be uniform. Kurtosis (Eq. \ref{eqn:kurtosis}) is defined as the ratio of the fourth moment (equation \ref{eqn:rth_moment} with $r = 4$) by the square of the variance (also equation \ref{eqn:rth_moment} with $r = 2$) \cite{zwillinger1999crc}

\begin{equation}
\label{eqn:rth_moment}
m_{r} = \frac{1}{n}
        \sum_{i=1}^{k}p_{i}(x_{i} - \bar{x})^{r}
\end{equation}

\begin{equation}
\label{eqn:kurtosis}
g_{2} = \frac{m_{4}}{(m_{2})^{2}} - 3.
\end{equation}

\noindent The $-3$ constant is due to Fischer's approach, where the kurtosis of a normal distribution is zero. Algorithm \ref{alg:kurtosis_array} denotes the pre-processing step:

\begin{algorithm}
	\caption{Kurtosis computation}
	\label{alg:kurtosis_array}
	\begin{algorithmic}[1]
	    \State // $X_{c \times n}$: dataset of $n$ descriptors with size $c \in C$, where \\ // $C = \{0,1,...,size(descriptor)\}$ 
	    
	    \\
	    
	    \State // $T(X)$: linear operator from equation \ref{eqn:probability_operator} to map the \\ // descriptors onto probability distributions
	   
        \\
        
		\State $X \gets T(X)$
		\State $A \gets zeros(c, n)$
		
		\For {\textbf{each} crop size $c$ in $C$}
		    \For{\textbf{each} descriptor $i$ in $\{1,2,...,n\}$}
		
        		\State $A[crop][i] \gets$ \textbf{kurtosis}$\left(X[i].subset(0, crop)\right)$
		
		    \EndFor
		\EndFor
		
		\\
		
		\Return $A$
	\end{algorithmic}
\end{algorithm}

\noindent Next, we propose a procedure to compute the optimal threshold. It is described in algorithm \ref{alg:cut_threshold}. The best crop size will be chosen such that the range within the dataset values is maximum. This will allow high-frequency information to be discarded from the  point spread function without loss of
blur information.

\begin{algorithm}[ht]
	\caption{Find the optimal dataset variability threshold}
	\label{alg:cut_threshold}
	\begin{algorithmic}[1]
 		\State // $A_{c \times n}$: matrix with kurtosis values for all $n$ descriptors // that were computed at every crop size $c \in C$, where \\ // $C = \{0,1,...,size(descriptor)\}$ 
		
		\\
		
		\State $threshold \gets 0$
		\State $maximum \gets \infty$
		\For {\textbf{each} crop size $c$ in $C$}
		\State $row \gets \{A_{c,1},A_{c,2},...,A_{c,n}\}$ 
		
		\State $a \gets$ \textbf{max}$(row)$
		
		\State $b \gets$ \textbf{min}$(row)$
		
		\\
		
		\If{$a < 0$ or $b < 0$}
		    \State \textbf{continue}
		\EndIf
		
		\\
		
		\State $range \gets a - b$
		
		\\
		
		\If{$range > maximum$}
		    \State $maximum \gets range$
		    \State $threshold \gets c$
		\EndIf
		
		\EndFor
		\\
		\Return $threshold$
	\end{algorithmic}
\end{algorithm}

With the optimal threshold, we can crop each descriptor without losing important frequency information. For each resultant vector, we
compute the interquartile range (IQR), which gives a measure of the spread for any distribution, whether or not it has a mean or variance. In a nutshell, it is the length of the interval that contains the middle half of the distribution
\cite{degroot2012probability}. Mathematically, it is the difference between the third ($Q_{3}$) and the first ($Q_{1}$) quartiles - values that separates the lowest $25\%$ of data from the highest $75\%$ and the highest $25\%$ of data from the lowest $75\%$, respectively \cite{devore2015probability}. IQR is described by

\begin{equation}
\label{eqn:iqr}
IQR = Q_{3} - Q_{1}.
\end{equation}

\noindent Next, for each distribution, we have a measure of its variability, which represents the sharpness metric of the corresponding image. In the scope of this work, higher variability means a higher amount of details (high frequencies) in the image. Therefore, the images with a higher interquartile range value should be classified as relatively sharp in accordance with a threshold.

This is achieved through another transformation, named $z$-score, that describes the location of an observation relative to the mean in units of the standard deviation. Given an arbitrary element $x$ of the distribution, a negative $z$-score shows that $x$ lies to the left of the mean, while a positive $z$-score indicate that the $y$ lies to the right of the mean \cite{mendenhall2016statistics}. The $z$-score is given by

\begin{equation}
\label{eqn:z-score}
z = \frac{x - \mu}{\sigma},
\end{equation}

\noindent where $\mu$ is the mean and $\sigma$ is the standard deviation of the distribution. Finally, the threshold to classify the images as sharp or blurred comes from the $z$-score, which measures how far an observation is from the mean of the dataset in terms of standard deviation units.

%% file: sections/experiments.tex
In this section, we describe the experimental setup for the evaluation of the proposed method. All tests have been conducted on an Intel Core i7 CPU computer with 8 GB RAM, running Ubuntu Linux 18.04 64-bit.

\subsection{Image dataset}

\begin{figure}
	\centering
	\includegraphics[scale=0.3]{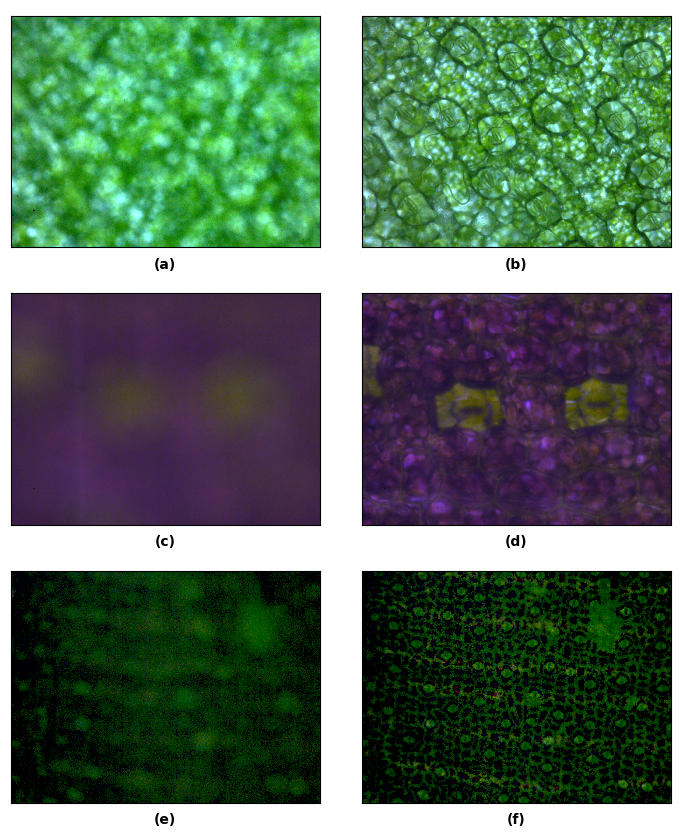}
	\caption{Examples of the proposed dataset images: blurred \textit{Callisia} \textbf{(a)}, sharp \textit{Callisia} \textbf{(b)}, blurred \textit{Tradescantia} \textbf{(c)}, sharp \textit{Tradescantia} \textbf{(d)} and blurred \textit{Cthenante} \textbf{(e)}, sharp \textit{Cthenante} \textbf{(f)}.}
	\label{fig:datasets}
\end{figure}

Three light microscopy datasets were acquired with the ZEISS SteREO
Discovery.v20 and the ZEISS AxioLab A1 stereo microscopes from the Scientific
Computing Group (SCG) at São Carlos Institute of Physics (IFSC). Samples of blurred and sharp images of both datasets are shown in Figure \ref{fig:datasets}.

The datasets contain images from leaf histological samples of the plants \emph{Callisia repens}, \emph{Tradescantia zebrina} and \textit{Cthenante oppenheimiana}, acquired with different focal planes and with different magnification levels.
In order to validate the results with correlation to a subjective quality index, each image was labeled as sharp or blurred, which respectively translates to \emph{eligible} and \emph{negligible} for the fusion process. The relevant properties of the datasets are summarized in Table \ref{tab:dataset_info}. 

\begin{table}[ht]
    \centering
    \begin{tabular}{lcccc}
        \toprule
        Dataset & Images & Mag. & Sharp & Sequence\\
        \midrule
        \textit{Callisia} & 56 & 50 & 9 & 41 - 49\\
        \textit{Tradescantia} & 66 & 200 & 2 & 50 - 51\\
        \textit{Cthenante} & 55 & 100 & 16 & 30 - 45\\
        \bottomrule
    \end{tabular}
    \caption{Information about the proposed datasets.}
    \label{tab:dataset_info}
\end{table}

\subsection{Validation metrics}

Three objective metrics were chosen to evaluate the performance of the proposed method. The classification of the images on our datasets can be considered as a subjective quality score, and therefore the objective metrics for comparison should relate to it. According to \citet{wang2011information}, evaluation metrics such as the Pearson Linear Correlation Coefficient (PLCC), the Spearman's Rank Correlation Coefficient (SRCC) and the Kendall's Rank Correlation Coefficient (KRCC) are suitable for the case. For all correlation coefficients, higher values yield higher reliability to the objective IQA metric.

%% file: sections/results-and-discussion.tex
We compared our proposed method with well-known NR-IQA approaches such as MLV \cite{bahrami2014fast}, $S_{3}$ \cite{vu2012s3}, JNB \cite{ferzli2009noreference}, CPBD \cite{narvekar2011noreference}, Marziliano \textit{et. al.} \cite{marziliano2002noreference} and Kanjar \cite{kanjar2013image}. Table \ref{tab:performance_comparison} shows the results of the PLCC, SRCC and KRCC metrics for the \textit{Callisia}, \textit{Tradescantia} and \textit{Cthenante} datasets, respectively.

\begin{table}[ht]
    \centering
    \begin{tabular}{lcccc}
        \toprule
        Dataset & Method & PLCC & SRCC & KRCC\\
        \midrule
        
        \multirow{7}{*}{\textit{\small Callisia}} 
        & MLV & 0.2829 & 0.2752 & 0.2267\\
        & $S_{3}$ & 0.1752 & 0.1730 & 0.1425\\
        & JNB & 0.5461 & 0.6031 & 0.4968\\
        & CPBD & 0.7361 & 0.6122 & 0.5043\\
        & Marz. & 0.7457 & 0.6122 & 0.5043\\
        & Kanjar & 0.6688 & 0.5971 & 0.4919\\
        & \textbf{Proposed} & \textbf{0.7488} & \textbf{0.6212} & \textbf{0.5117}\\
        
        \midrule
        
        \multirow{7}{*}{\textit{\small Tradescantia}} 
        & MLV & 0.1285 & 0.1346 & 0.1107\\
        & $S_{3}$ & -0.1312 & -0.1253 & -0.1031\\
        & JNB & 0.2561 & 0.2181 & 0.1794\\
        & CPBD & 0.2409 & 0.2413 & 0.1985\\
        & Marz. & 0.2322 & 0.2227 & 0.1832\\
        & Kanjar & 0.2564 & 0.2227 & 0.1833\\
        & \textbf{Proposed} & \textbf{0.3698} & \textbf{0.2552} & \textbf{0.2099}\\

        \midrule
        
        \multirow{7}{*}{\textit{\small Cthenante}} 
        & MLV & 0.0446 & 0.0227 & 0.0187\\
        & $S_{3}$ & -0.1682 & -0.2068 & -0.1704\\
        & JNB & 0.7840 & 0.7414 & 0.6108\\
        & CPBD & 0.8041 & \textbf{0.7515} & \textbf{0.6191}\\
        & Marz. & 0.8012 & 0.7464 & 0.6150\\
        & Kanjar & 0.7831 & 0.7338 & 0.6046\\
        & \textbf{Proposed} & \textbf{0.8129} & 0.7414 & 0.6108\\
        
        \bottomrule
    \end{tabular}
    \caption{Performance comparison of our proposed method and other NR-IQA metrics on the microscopy images datasets.}
    \label{tab:performance_comparison}
\end{table}

From Table \ref{tab:performance_comparison}, we may conclude that the proposed method produced reasonable results when applied to the proposed microscopy images. The difference between our images and benchmark ones such as LIVE \cite{sheikh2006statistical} and CSIQ \cite{larson2010most} is that the light microscopy images are subjected to a non-homogeneous blur kernel; there are spherical aberrations, i.e. natural deformations of the lenses' surface that promotes scattering of the refracted light rays. Therefore, the classification of each image is even more subjective, since the notion of quality might be different according to what the images will be used for. In the scope of this work, the reason for assessing image quality is to perform a fusion process which is our final goal. This depends on an analysis of the predicted eligible images. 

Another relevant feature concerning any NR-IQA index is the monotonicity. Considering the set which represents the values obtained by applying the metric on an image dataset and the set of labels provided by the subject evaluation of the dataset, the monotonicity is the property of maintaining the order relation between the sets, i.e. it is only nondecreasing or nonincreasing. Suitably, the chosen correlation coefficients to evaluate the performance of the metric also denote monotonicity. The results in Table \ref{tab:performance_comparison} quantitatively show how monotonic the metric is.

Finally, the computational performance of an NR-IQA \linebreak metric may be a constraint. As an example, if the metric is employed in microscope auto-focus systems, the time to yield the results should be minimized, considering that the computations will be updated several times until the sharpest focus configuration is found. The size of the microscopy images also implies an execution time limitation. Our method was built with the C++ language and the OpenCV \footnote{https://opencv.org/} framework in order to achieve good computational performance with the hardware and operating system settings shown in section \ref{sec:experiments}, and is available in a repository\footnote{https://github.com/vaugusto92/fourier-light-microscopy-nr-ism}. We implemented the Kanjar method in Python, and the code is also organized in a repository\footnote{https://github.com/vaugusto92/kanjar-nr-iqa}. The other NR-IQA methods described in section \ref{sec:related-work} were implemented in MATLAB, C++ or Python programming languages, and Table \ref{tab:running_times_comparison} presents the running time comparison for all methods, when applied in our proposed \textit{Callisia repens} dataset. Despite the programming language differences, our method yielded a relevant computational efficiency in terms of execution time.

\begin{table}[ht]
    \centering
    \begin{tabular}{lcc}
        \toprule
        Method & Frameworks & Time\\
        \midrule
        
        MLV & MATLAB\tablefootnote{https://sites.google.com/site/khosrobahrami2010/publications} &
        1 minute and 6 seconds\\
        
        $S_{3}$ &
        MATLAB\tablefootnote{http://vision.eng.shizuoka.ac.jp/s3/} &
        4 hours and 25 minutes\\
        
        JNB & 
        MATLAB\tablefootnote{https://ivulab.asu.edu/software/jnbm/} &
        4 minutes and 37 seconds\\
        
        CPBD & Python, cpbd\tablefootnote{https://pypi.org/project/cpbd/} &
        14 minutes\\
        
        Marz. & C++ \tablefootnote{https://github.com/PeterWang1986/blur}, OpenCV &
        8 seconds\\
        
        Kanjar & 
        Python, NumPy &
        25 seconds\\
        
        Proposed & 
        C++, OpenCV &
        13 seconds\\
        
        \bottomrule
    \end{tabular}
    \caption{Running times for the proposed method and other NR-IQA metrics on our \emph{Callisia repens} microscopy images dataset.}
    \label{tab:running_times_comparison}
\end{table}

%% file: sections/conclusions.tex
We proposed a no-reference Discrete Fourier Transform-based image quality metric. It produces results that highly correlate with the labels obtained by subjective analysis. The implementation of the method is efficient in terms of computational performance, and improving this performance is one of the aims of future work. The main application is providing a reliable representation of the quality of each image in order to select the ones with sharp regions and perform a fusion process, which will result in a good quality image. It can also be used in auto-focus systems for capturing devices (not only the microscope). Further research improvements are optimization of the implementation, integration with hardware devices and the development of other classification algorithms which may perform better than the $z$-score, which may include more advanced machine learning tools.